\documentclass[12pt]{article}
\pdfoutput=1
\usepackage{graphicx}
\usepackage{color}
\usepackage{cancel}
\usepackage{comment}

\newcommand{\lsim}{\raisebox{-0.13cm}{~\shortstack{$<$ \\[-0.07cm] $\sim$}}~} 
\newcommand{\gsim}{\raisebox{-0.13cm}{~\shortstack{$>$ \\[-0.07cm] $\sim$}}~} 
\newcommand{\beq}{\begin{eqnarray}} 
\newcommand{\eeq}{\end{eqnarray}} 
\newcommand{\tb}{\tan\beta}

\newcommand{\bea}{\begin{align}}
\newcommand{\eea}{\end{align}}
\newcommand{\nbea}{\begin{align*}}
\newcommand{\neea}{\end{align*}}
\newcommand{\nbeq}{\begin{equation*}}
\newcommand{\neeq}{\end{equation*}}
\newcommand{\bear}{\begin{eqnarray}}  
\newcommand{\eear}{\end{eqnarray}}  

\begin{document}

\begin{flushright}
KCL-PH-TH/2024-77, CERN-TH-2024-219
\end{flushright} 

\vspace*{5mm} 

\begin{center}

\mbox{\large\bf Contrasting {Pseudoscalar Higgs and Toponium}}

\vspace*{2mm}

\mbox{\large\bf States at the LHC and Beyond}

\vspace*{8mm}

{\sc Abdelhak~Djouadi$^{1,2}$, John Ellis$^{3,4}$} and {\sc J\'er\'emie Quevillon$^1$} 

\vspace{6mm}

{\small 
$^1$ Laboratoire d'Annecy de Physique Th\'eorique, 9 Chemin de Bellevue, 74940 Annecy, France.\\[2mm]

\mbox{ \hspace*{-6mm} $^2$ Departamento de F\'isica Te\'orica y del Cosmos, Universidad de Granada,
18071 Granada, Spain.} \\[2mm] 

$^3$ Physics Department, King’s College London, Strand, London, WC2R 2LS, United Kingdom. \\[2mm] 

$^4$ Theoretical Physics Department, CERN, CH 1211 Geneva, Switzerland. 
}

\end{center}

\vspace*{8mm}

\begin{abstract}
We discuss the phenomenology at hadron colliders of a pseudoscalar Higgs boson $A$, as predicted in many extensions of the Standard Model. In particular, we highlight the aspects that makes it different from the case of a quasi-bound toponium state. We apply the discussion to the excess of $t \bar t$ threshold events recently observed at the LHC by the CMS collaboration \cite{CMS},  which could in principle be due to either or both possibilities, although Occam's razor favours the toponium scenario. Working in an effective theory in which only an additional pseudoscalar $A$ boson is present in the spectrum, with a mass above  the $2 m_t$ threshold and a significant coupling to top quarks, we discuss  the interference between $A$ production in the dominant gluon-fusion process $gg \to A$ with subsequent $A\to t\bar t$ decays, and the QCD continuum background, $gg\to t \bar t $. While this interference is absent in the case of toponium, it is essential for evaluating $A$ production.  It is difficult to resolve the peak/dip structure that it generates because of the experimental smearing of the $t\bar t$ invariant mass spectrum. {However, by comparing the total $A$ production rates for different integration domains of the $t\bar t$ invariant mass or, eventually, at different center-of-mass energies, one may be able to observe its effects}.  We then discuss additional mechanisms for $A$ production in $pp$ collisions, including loop-induced production in association with the lighter $h$ boson, $gg \to hA$, and production in association with top-quark pairs, $gg/q\bar q \to t\bar t A$.  These mechanisms have small cross sections at the LHC, and their observation will necessitate higher luminosities or collider energies.\\
~~\\
{\it We dedicate this paper to the memory of our dear friend Rohini Godbole, with whom we performed a study along similar lines \cite{Rohini}.}\\
\end{abstract}
\vspace*{1cm}

\newpage

\subsection*{1. Introduction}

Searches for additional Higgs bosons rank high among probes of possible new physics beyond the Standard Model (SM), and feature strongly in the research programmes of the CMS and ATLAS  experiments at the LHC and in studies for proposed next-generation particle colliders.  The most recent such search by the CMS collaboration~\cite{CMS} has produced an intriguing result, namely a 5$\sigma$ enhancement of events close to the $t \bar t$ threshold containing final states with leptons and jets, whose angular correlations are characteristic of production of a pseudoscalar spin-parity state. This signal has yet to be confirmed by the ATLAS collaboration, but one may nevertheless start asking whether it could be due to a new pseudoscalar Higgs boson $A$ that occurs in extensions of the SM, or if it is a QCD effect. The $A$ hypothesis is particularly relevant if the mass of the resonance is significantly different from twice the top quark mass. 

The primary candidate for the CMS signal, and the one that Occam's razor favours, is \underline{clearly} a $^1S_0$ quasi-bound $t \bar t$ state, referred to here as `toponium’ and denoted by $\eta_t$. It is a firm prediction of QCD that there should be some degree of Sommerfeld (Coulomb) enhancement in the $^1S_0$ channel around the $t \bar t$ threshold, and there have been detailed studies of the analogous effect in the $^3S_1$ channel and the observability of vector toponium in $e^+ e^-$ collisions~\cite{Toponium,Reuter}. Calculations of the $^1S_0$ enhancement at the LHC are much less advanced but the estimates of the $\eta_t$ production cross section at the LHC at at a c.m. energy of 13~TeV ~\cite{Kiyo, Sumino, Fuks} seem to be consistent with the strength of the CMS signal.~\footnote{There has been a long debate within the LHC community whether there is a physical pseudoscalar toponium state, since its $t$  and $\bar t$ constituents could decay before they could bind; for a recent discussion of the Sommerfeld enhancement for unstable particles, see Ref.~\cite{Luo}. Here, we set aside this discussion and use the word `toponium’ as shorthand for a QCD enhancement in the pseudoscalar $t \bar t$ channel.}  

A second option to explain the observed excess of events, which was considered by the CMS collaboration itself in Ref.~\cite{CMS},
is the presence  of a new pseudoscalar Higgs boson $A$ in the spectrum, with a mass of $M_A=365$ GeV and large couplings to the top quarks, in addition to the 125 GeV scalar $h$ state already observed at the LHC \cite{Hdiscovery}.  It is important to 
be able to distinguish between the signatures of the $\eta_t$ and $A$ scenarios, in particular if the  two states are very close in mass (which seems not to be the case here). As discussed in Ref.~\cite{CMS}, the $\eta_t$ and $A$ hypotheses make identical predictions for the final-state angular correlations between the leptons and jets produced in $t$ and $\bar t$ decays, as these are determined by the spin correlations arising from the decay of a CP--odd state~\cite{Afik,Maltoni}. 

However, as also discussed in Ref.~\cite{CMS}, the $A$ hypothesis offers the possibility of interference between the production process $gg\to A$ with the subsequent decay $A\to t\bar t$ and the QCD continuum ``background" $gg \to t\bar t$ as both have the same initial- and final-state topologies~\cite{Nous}. CMS discussed this interference \cite{CMS}, making {\it ad hoc} assumptions about the decay width of the $A$ boson.  In turn, there is no such possibility under the toponium hypothesis, as the resonant signal is due to the QCD  ``background" itself. 

In this paper, we revisit the impact of the interference of a resonance signal with the QCD background in $t\bar t$ final states in an effective scenario in which a new pseudoscalar Higgs boson $A$, with a mass of about 365 GeV and couplings to top quarks that could lead to the signal advocated by CMS as a second possibility, 
is present in the spectrum~\footnote{The $A$ state will nevertheless have properties that are similar to that of the pseudoscalar Higgs boson that occurs in the context of realistic extensions of the scalar sector in which other Higgs states are present, such as the two-Higgs doublet model (2HDM) \cite{2HDM}, possibly with an additional light pseudoscalar state (2HD+a) \cite{2HDa}.  The 2HDM option was discussed in Ref.~\cite{Cheung} in the context of the CMS signal, and we compare some of our results with theirs below.}.  The $A$ state is assumed to couple most strongly to top quarks and be produced primarily in the gluon-gluon fusion process, $gg\to A$, which is mediated by triangular top quark loops. For the mass and couplings that yield the appropriate normalisation of the CMS signal, the decay width of the $A$ boson is almost fixed, constraining the possible magnitude of the interference between the $A$ signal and the QCD background.  

{There are both real and imaginary contributions to the interference: the former changes sign across the resonance peak, whereas the latter is negative~\cite{Nous}. The interference has a significant impact on the total $A$ production cross section, with the negative real interference extending to values of $m_{t \bar t}$ well beyond the nominal resonance width. As a result of the poor experimental resolution on the $t\bar t$ invariant mass of $\approx 20\%$ as reported by the CMS collaboration, the interference is difficult to observe experimentally. Nevertheless, interference effects could be probed by precise measurements of $\sigma(t \bar t)$ over a wide range of $m_{t \bar t}$ values, or by measuring the $gg\to A \to t\bar t$ cross section at different collider energies, as we discuss below.} 

As the $A$ branching ratios for decays into final states such as $\gamma\gamma, \gamma Z, ZZ,WW$ could be the same as those of toponium, one should consider other means to distinguish $A$ from it and consider $A$ production by other mechanisms at the LHC and, eventually, at higher-energy $pp$ colliders as well as signatures in $e^+ e^-$ collisions. Here, we study two processes in which a relatively light $A$ boson can be produced at hadron colliders apart from $gg \to A$, namely, associated production of $h$ and $A$ through box diagrams involving top quarks in gluon fusion, $gg\to hA$, and radiation from heavy top quark pairs, $gg, q\bar q \to t\bar tA$. We show that in both cases the rates are very small at $\sqrt s=13$ TeV for parameters consistent with the CMS signal, whereas a ${\cal O}(100)$~TeV collider would have a much large event sample that could allow for a clear observation. We comment briefly on the relevant signals in $e^+e^-$ collisions and on possible production and detection mechanisms in model extensions in which there are additional neutral and charged Higgs particles in the spectrum. 

The rest of the paper is organized as follows. In Section~2, we present the salient features of a pseudoscalar $A$ state with a mass of $\sim 365$ GeV and couplings to top quarks that reproduce the CMS signal, in  the context of an effective theory. In Section~3, we discuss the impact of the interference effects between the $A$ resonance signal and the QCD $t\bar t$ background and how it can be probed.  In Section~4, we study the other processes in which $A$ can be produced at hadron colliders. Finally, Section 5 summarises our conclusions.

\subsection*{2. Phenomenology of a 365~GeV Pseudoscalar Boson}

Following Ref.~\cite{Rohini}, and staying as model-independent as possible, we simply introduce an isospin-singlet pseudoscalar $A$ boson that has an effective interaction only with the heavy top quarks, neglecting all couplings with other fermions, including $b$ quarks and $\tau$ leptons. The interaction Lagrangian, given by $\mathcal{L}_{\rm Yuk} \supset i \hat g_{Att}  \bar{t} \gamma_5 t A$, can be generated, for instance, via an effective operator of dimension 5 or higher. Using the SM Higgs Yukawa interaction with fermions as a reference,  the new $At\bar t$ coupling can be written in terms of the quark mass and the SM vacuum expectation value $v$ as  
 \begin{eqnarray} 
\hat g_{A t t}= \frac{m_t}{v} \times g_{A t t}  \quad {\rm with}~v=246~{\rm GeV} \, . 
 \label{eq:yukawa}
\end{eqnarray} 
In the context of a two-Higgs doublet model (2HDM) \cite{2HDM} with neutral fields  $\Phi_1$ and $\Phi_2$ developing vacuum expectation values $v_1$ and $v_2$, where $\sqrt {v_1^2+v_2^2} = v$ and with $\tb \equiv v_2/v_1$, the reduced Yukawa coupling above is uniquely fixed in term of the ratio of vevs, $g_{Att}= \cot \beta $. For small values $\tan\beta \lsim 1$, this coupling to top quarks is rather strong, while the $A$ couplings to all other fermions are suppressed by the much smaller quark or lepton  masses and not strongly enhanced by $\tan\beta$ factors.

The Higgs interactions with fermions in a 2HDM are model--dependent, with two major options usually discussed~\cite{2HDM}: in the Type-II scenario, the field $\Phi_1$ ($\Phi_2$) generates the isospin down--type fermion (up-type quark) masses, whereas in the Type-I scenario $\Phi_2$ couples to both up-- and down--type fermions. One finds for the couplings to third-generation fermions:  $g_{Att}\!=\! \cot \beta$ in both Type-I and II and $g_{Abb}\!=\!g_{A\tau\tau}\!=\! \cot\beta$ in Type-I while $g_{Abb}\!=\!g_{A\tau\tau}\!=\! \tan\beta$ in Type-II. Two more options, called Type-X and Type-Y in which the couplings of leptons and down-type quarks are different are also discussed~\cite{2HDM}. In all cases, one can neglect all couplings expect for $\hat g_{Att}$ when $\tan\beta \approx 1$. 

Hence, the situation in a 2HDM coincides with the effective one that we are considering here. More generally, any 2HDM, which has a spectrum of two CP--even neutral bosons $h$ and $H$  and two charged $H^\pm$ bosons in addition to the CP--odd boson $A$, can  reduce to the isosinglet $A$ scenario that we are considering. This occurs if two conditions, which automatically satisfy all constraints from LHC and other collider data, are met:  $a)$ one is in the so-called alignment limit~\cite{alignment} in which the lightest $h$ is identified with the observed 125 GeV Higgs boson and has exactly SM-like couplings to fermions and gauge bosons (we note that there is no $AhZ$ coupling in this case, and that the $AZZ$ and $AWW$ couplings are forbidden by CP invariance); and $b)$ the heavier neutral $H$ and charged $H^\pm$ bosons  are assumed to be much heavier than $A$, $M_H \approx M_{H^\pm} \gg M_A$, and are not observable neither directly at the LHC \cite{HHunting,Htautau} nor indirectly in high precision measurements of electroweak observables \cite{Deltarho} or heavy flavors \cite{bsgamma}.

In our scenario, as there are no other Higgs bosons, there are possible self-couplings only among $h$ and $A$ states, which we can tune to be small and even set to zero. However, we recall that these self-couplings set strong theoretical constraints in specific models like the 2HDM (in particular from the requirement of perturbativity), and might even exclude in such models the possibility of a 365 GeV $A$ state, as  discussed in Ref.~\cite{Cheung} for instance. 

Once the mass $M_A$ and the reduced coupling $g_{Att}$ have been  chosen, the phenomenology of the $A$ state is fixed. The interactions of $A$ with massive gauge bosons, photons and gluons, as well as the $AhZ$ coupling, are induced at the one-loop level through triangular diagrams involving the contributions of the top quark with strength $g_{Att}$. Thus, the $A$ decay widths and the branching ratios are fixed, as also are the production cross sections.  

For instance, defining the reduced mass variable $\tau= M_A^2/4m_t^2$, the top-quark loop induced decay of the $A$ boson into two gluons is given by \cite{Anatomy2}
\begin{eqnarray}
\Gamma(A  \to gg) =   \frac{G_F \alpha_s^2 M_A^3} {16\sqrt{2}\pi^3} g^2_{A tt} \bigg|\frac{ f(\tau) }{\tau} \bigg|^2, ~f(\tau)=\left\{ \begin{array}{ll}  \displaystyle \arcsin^2 ({1}/{\sqrt{\tau}}) & {\rm for} \; \tau \geq 1 \, , \\
\displaystyle -\frac{1}{4}\left[ \log\frac{1+\sqrt{1-\tau }} {1-\sqrt{1-\tau}}-i\pi \right]^2 & {\rm for} \; \tau<1 \, .
\end{array} \right.
\end{eqnarray}
The hadron-level $A$ production cross section in the dominant gluon-gluon fusion process $gg\to A$  can be written at leading order in terms of this partial width:  
\begin{eqnarray}
\sigma(pp \rightarrow A) = \frac{1}{M_A s} C_{gg} \Gamma(A \rightarrow gg) \, :~
C_{gg} =  \frac{\pi^2}{8} \int_{M_A^2/s}^{1} \frac{dx}{x} g(x)g({M_A^2}/{s x} ) \, ,
\end{eqnarray}
where $g(x)$ is the gluon distribution inside the proton defined at a suitable factorization scale $\mu_F$, while the coupling $\alpha_s$ is evaluated at the renormalisation scale $\mu_R$. {As we are probing the structure of the resonance near the pole, we set  $\mu_F=\mu_R= M_A$.}   

We note also that, as we have assumed that the $A$ state is isosinglet, and also because of CP invariance that forbids $A$ boson couplings to massive $W^\pm$ or $Z$ bosons, its production in association with, or in the fusion of, massive gauge bosons are strongly suppressed. 

As is well known, one cannot be satisfied with this leading-order picture and should include higher-order effects, which are important in this case. In particular, the QCD corrections that increase the production rate by more than a factor of two need to be incorporated.  We calculate  the total $gg\to A$ cross section using the numerical program {\tt SUSHI} \cite{SUSHI}, which incorporates these higher-order QCD effects. When using this program, the $gg\to A$ process can be evaluated ignoring the bottom quark loop contribution which is absent in our scenario. In a 2HDM, this contribution might be significant if the $Abb$ coupling is strongly enhanced, but this is not the favored scenario in the case under discussion, so one obtains approximately the same results in general. The QCD corrections, which are known up to next-to-leading order in the massive case \cite{ANLO} and at next-to-next to-leading order in the effective limit $M_A\! \ll\! 2m_t$ \cite{ANNLO}, give a $K$-factor of about 2.

As for $A$ decays, the only possible mode that occurs at tree level  in our context is that into top quark pairs, $A \to t\bar t$.  If $A$ is sufficiently massive, $M_A \gsim 2m_t = 345$ GeV (we assume for definiteness a top quark mass value of $m_t=172.5$ GeV \cite{ATLAS:2024dxp} in our discussion) it will be a two-body decay mode. The partial decay width, as no other decay mode is present at tree level~\footnote{This is again also approximately the case in a 2HDM, since the decays $A \to b\bar b, \tau^+\tau^-$ are suppressed for $\tan\beta \lsim 1$ and the mode $A \to hZ$ is absent in the alignment limit in which the $AZh$ coupling vanishes. All other decay channels are absent if $A$ is lighter than the $H$ and $H^\pm$ particles.}  and loop processes are suppressed, is equivalent to the total decay width $\Gamma_A$ to a good approximation. At leading order, the two are given in terms of the top quark velocity $\beta_t=(1-4m_t^2/M_{A}^2)^{1/2}$  by \cite{Anatomy2} 
\begin{eqnarray}
\Gamma_A = \Gamma(A \to t \bar t) = 3 \frac{ G_F m_t^2}{4\sqrt{2} \pi} \,  g_{A tt}^2 \, M_{A} \, \beta_t \, .  \label{eq:Gamma}
\end{eqnarray}
Here again, higher-order effects should be implemented in general. Besides the QCD corrections, which are important also in this case, one needs to consider the possibility of three-body $A$ decays with one top quark  off mass-shell and decaying into a $Wb$ final state, $A \to t\bar t^* \to tbW$. This implies that for $M_A <2 m_t$, the total width, although rather small, is non-zero. However, the branching ratio is still close to unity for masses not much smaller than the $2m_t$ threshold.~\footnote{Note that this might not be case in a 2HDM, as decays into $b$ quarks and $\tau$ leptons are then possible and might dominate over the $A \to tt^*$ mode, especially if the couplings $g_{Abb}$ and $g_{A\tau\tau}$ are enhanced.}. For an accurate determination of the total width $\Gamma_A$, we use the program  {\tt HDECAY}~\cite{HDECAY}, which includes all the relevant higher-order effects. For a mass  $M_A$ only slightly  above the $t\bar t$ threshold and for a coupling $g_{Att}$ of order unity, one obtains a width to mass ratio $\Gamma_A/M_A \simeq 2\%$.  More  precisely, for $M_A=365$ GeV and $g_{Att}=1$, $\Gamma_A=10.7$ GeV for both the partial and total widths. {This should be contrasted with the value $\Gamma_A^0 \simeq 6.9$ GeV obtained in the Born approximation using Eq.~(\ref{eq:Gamma}).}

We see from this discussion that the total $A$ decay width is much smaller than the experimental resolution of the $t\bar t$ invariant mass in the CMS detector, which is estimated to be $\sim 20$\%~\cite{CMS}.  Hence the $A$ state is a rather narrow resonance in our case  and cannot be experimentally resolved. For comparison, we recall that the CMS analysis \cite{CMS} considered several {\it ad hoc} possibilities for $\Gamma_A^{\rm tot}/ M_A$ in the range (1--25)\%, with $\Gamma_A^{\rm tot}/ M_A = 2$\% being adopted as the default assumption in many figures and in Appendix~C.\vspace*{-2mm}

\subsection*{3. Interference effects in $\mathbf{gg\to A \to t\bar t}$} 

The contrasting behaviour between the production of a pseudoscalar boson $A$ with a mass around the $t\bar t$ threshold and a pseudoscalar color-singlet toponium quasi-bound state $\eta_t$ is in principle obvious. The latter  is made exclusively by the QCD process $gg \to t\bar t$ (at the LHC, the contribution of $q\bar q$ annihilation is rather small), whereas in the former, the final-state  is produced by both the QCD continuum $t \bar t$ production process and also the resonant electroweak process $gg \to A\to t\bar t$. In general, since these two processes share the same initial and final states, they will interfere. This leads not only to a change in the production rate but also to a distortion of the resonance line-shape.  This interference may be either constructive and/or destructive, depending on the kinematics, giving rise to a more complex signature with a peak–dip structure in the $t \bar t$ invariant mass distribution. 

Calculations of the cross section of the signal and continuum background processes, as well as their interference, have been performed {e.g. in Ref.~\cite{Nous}, which we follow closely here; other studies can be found  in Refs.~\cite{Autres}.}  At leading order, the full matrix elements including all mass effects are taken into account in calculating the signal, background and interference parts. Higher-order QCD corrections are implemented using suitable short-cuts and some widely-used approximations that simplify the calculation but lead to rather accurate predictions, at least for the known cases of signal and background.

We use here the same approximations for the higher-order effects as {in Refs.~\cite{Maltoni,Nous} and} in the CMS analysis~\cite{CMS}. For the Higgs signal, the K--factor of order 2 obtained at NNLO \cite{ANLO,ANNLO}  is adopted, while for the background we assume the K--factor $\simeq 1.3$ found at NNLO~\cite{QCDtt}; in both cases we ignore the small electroweak corrections. For the interference part, the higher-order QCD corrections are {not} known and, following CMS~\cite{CMS} and Refs.~\cite{Maltoni,Nous}, we use as a K--factor the geometrical average of the signal and QCD background {K--factors at NNLO, $K \simeq 1.6$. 

For consistency,  the total $A$ decay width that should appear in the interference term is the one calculated at leading order in perturbation theory.  For $M_A\!=\!365$ GeV and $g_{Att}\!=\!0.78$, it  has the value $\Gamma_A^{\rm LO} \! \simeq \! 4.2$ GeV, instead of the value $\Gamma_A^{\rm HO} \! \simeq \! 6.5$ GeV  that higher-order QCD effects, which are, in principle, taken into account via the K-factor and is obtained, e.g., using the program {\tt HDECAY}. It turns out that the ratio of widths,  $\Gamma_A^{\rm HO}/\Gamma_A^{\rm LO} \simeq 1.6$, is approximately the same as the assumed QCD K--factor in the interference.}
We also note that for $M_A \gsim 350$ GeV the fraction BR($A \to t\bar t)$ is very close to unity in all cases, except very close to the $2m_t$ threshold where it is only very slightly suppressed. {However, we stay away from the threshold region $\lsim 2m_t$.}

{We focus our numerical analysis on a pseudoscalar Higgs boson with a mass of $M_A=365$ GeV and a strong coupling to the top quark, $g_{Att} =0.78$, which is the best-fit point of the CMS collaboration for this particular scenario. We are, therefore, well above the $t\bar t$ threshold $m_{tt}=345$ GeV and we do not expect any effect from a toponium bound state in the production process nor large corrections to the branching ratio in the decay mode $A \to t\bar t$. We therefore believe that our analysis is reliable and not affected by the non-perturbative threshold effects \cite{ttthreshold}.

Nevertheless, \underline{for the sake of illustration}, we also mention one example in which the Higgs mass is smaller, $M_A=350$ GeV, just to see how the interference behaves. For such a value, the $A$ total decay width is $\Gamma_A \simeq 2.3$ GeV so that $M_A \gsim m_{tt} + 2\Gamma_A$, which should, hopefully, also relatively safe from non-perturbative effects.} \footnote{According to Ref.~\cite{Fuks}, the toponium contribution to the cross section is important only below the $2m_t \approx 345$ GeV threshold; this is also apparent in the earlier paper Ref.~\cite{Kiyo}. For the decay, the mode $A\to t \bar t$ is by far dominant and its branching ratio is very close to unity, and hardly affected by non-perturbative or higher order effects.}  We will also not enter into experimental issues such as the ones discussed, e.g., in the first paper in Ref.~\cite{Nous}, and simply show the outcome of a simple parton-level analysis that nevertheless captures the main relevant aspects and provides a reliable assessment of the situation. 

\begin{figure}[!ht]
\vspace*{-5mm}
\begin{center}  
\mbox{\includegraphics[scale=0.34]{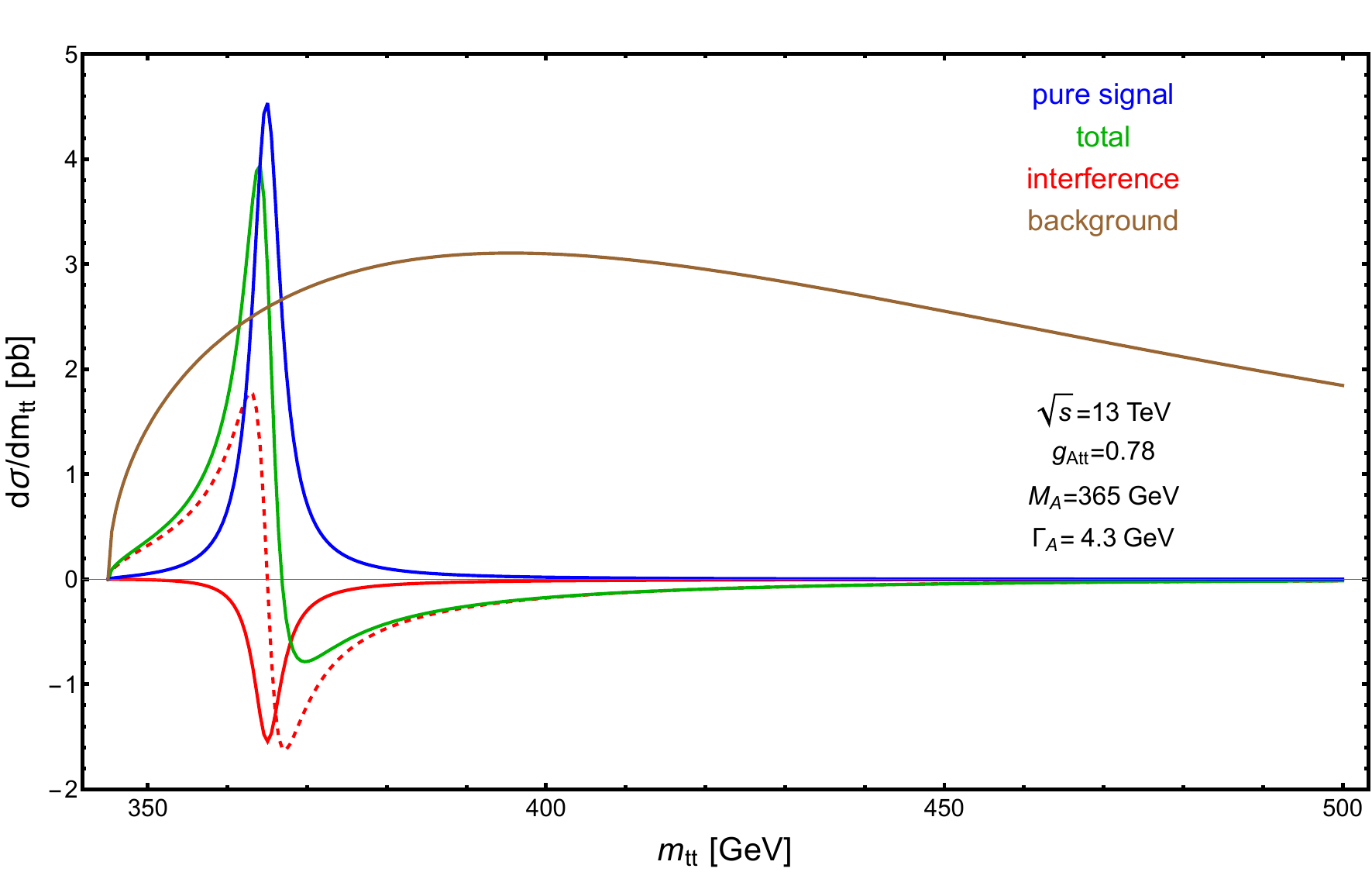}
}
\end{center} 
\vspace*{-6mm}
\caption{\it The contributions to the line–shape of a pseudoscalar $A$ state with a mass $M_A\!=\!365$ GeV in the process $gg\! \to \! A \! \to \! t \bar t$ at  the $\sqrt s\!=\!13$ TeV LHC. We show the contributions from the pure signal only (blue line), the continuum $t \bar t$ background (brown line), the real and imaginary interference contributions (dashed and solid red lines) and the total cross section including the interference (green line). Also shown is the value of the $g_{Att}$ coupling that is determined by the rate measured by CMS~\cite{CMS}, and the resulting value of $\Gamma_A$.}
\label{fig:365}
\vspace*{-4mm}
\end{figure}

{Fig.~\ref{fig:365} displays the theoretical line-shape for an $A$ boson with the production cross section reported by CMS~\cite{CMS}, namely 365~GeV. It includes the pure signal (blue line), the continuum $t \bar t$ background (brown line), the real and imaginary contributions to the interference (dashed and solid red lines, respectively) and the total cross section (green line).
{We see that the (negative) contribution due to the real part of the interference is significant out to values of $m_{t \bar t}$ far beyond the natural Breit-Wigner width of the $A$ boson.} We comment that the value of the $g_{Att}$ coupling and hence the total $A$ width is determined by its mass and the cross section reported by CMS~\cite{CMS}. Since the continuum $t \bar t$ background rises rapidly above the nominal $t \bar t$ threshold, the magnitudes of the interference effects also increase with $m_{t \bar t}$, and also their importance relative to the pure signal peak.

\begin{figure}[!ht]
\vspace*{-4mm}
\begin{center}  
\mbox{\includegraphics[scale=0.32]{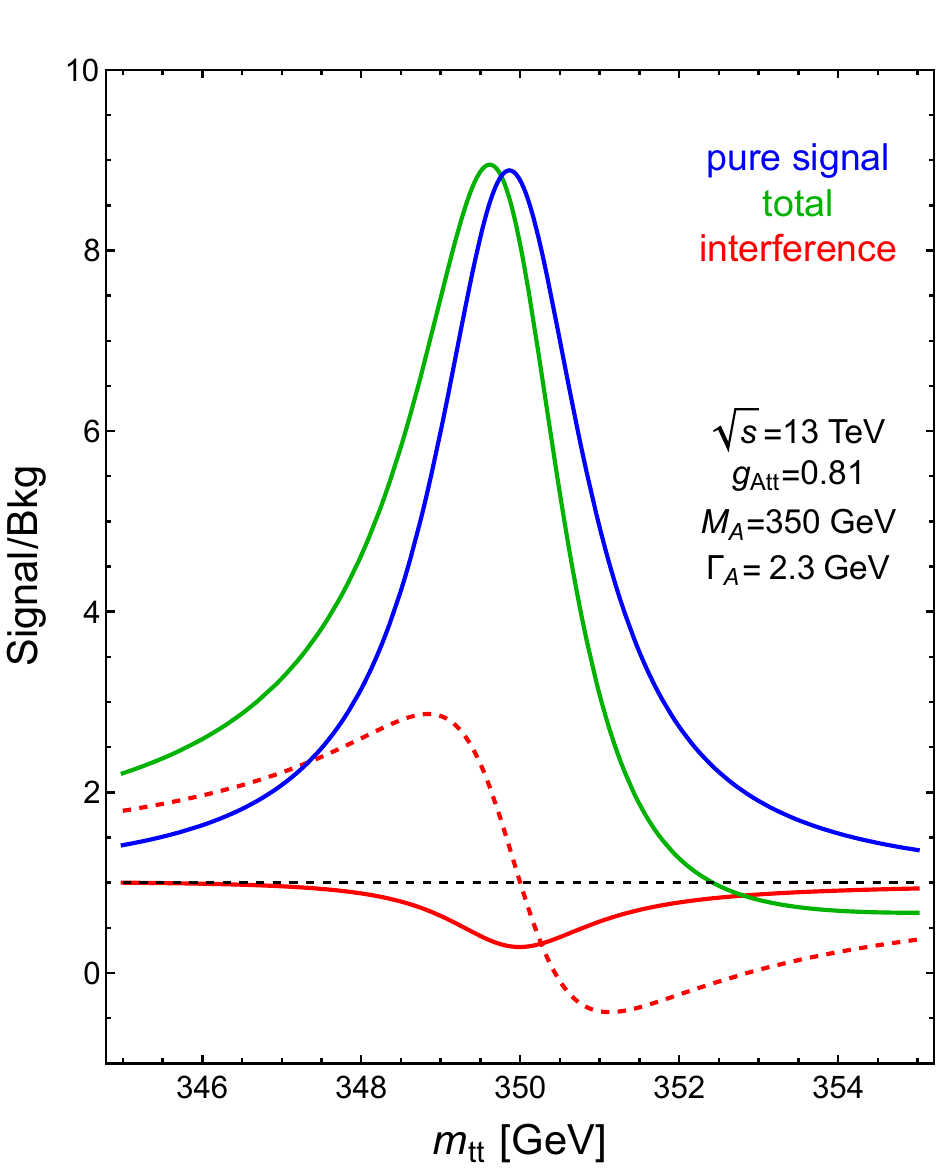}~ 
      \includegraphics[scale=0.32]{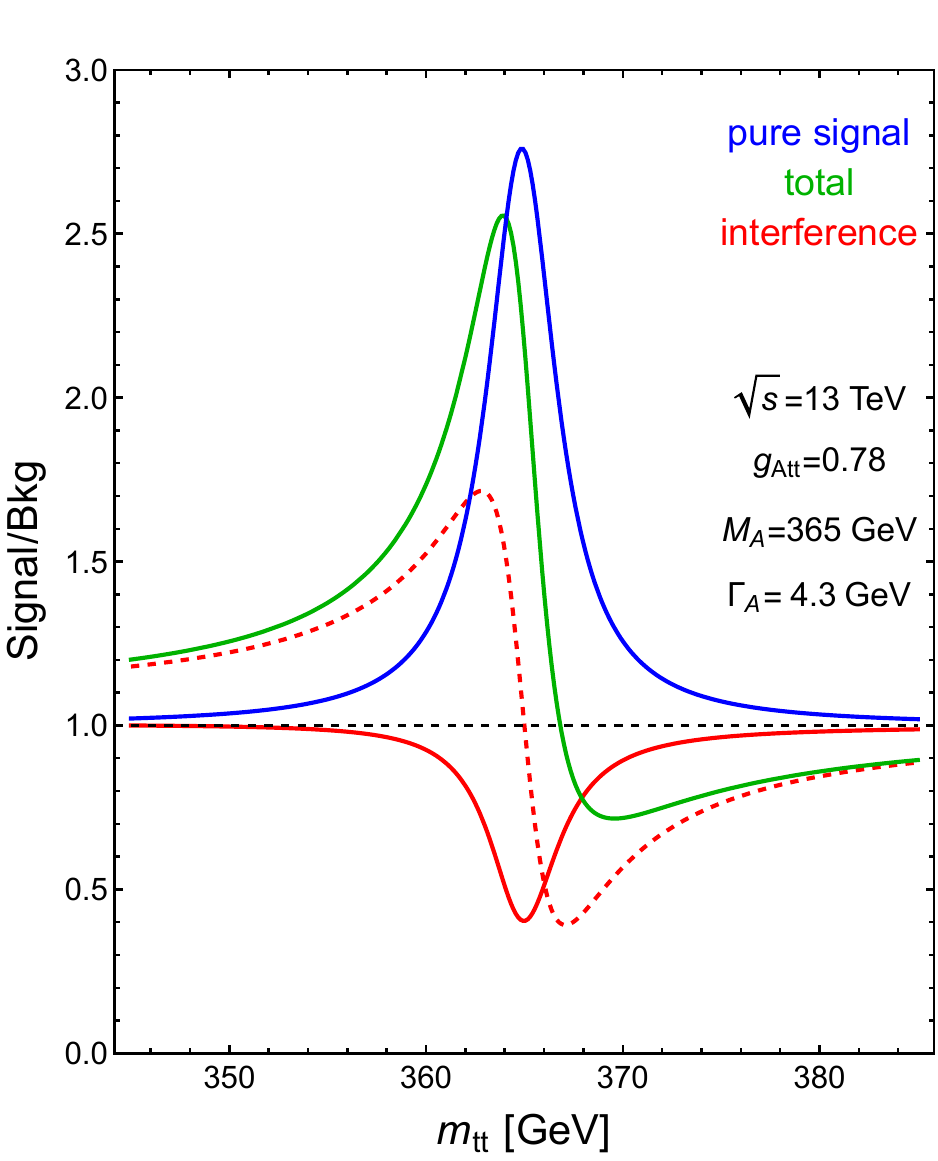}~ 
      \includegraphics[scale=0.32]{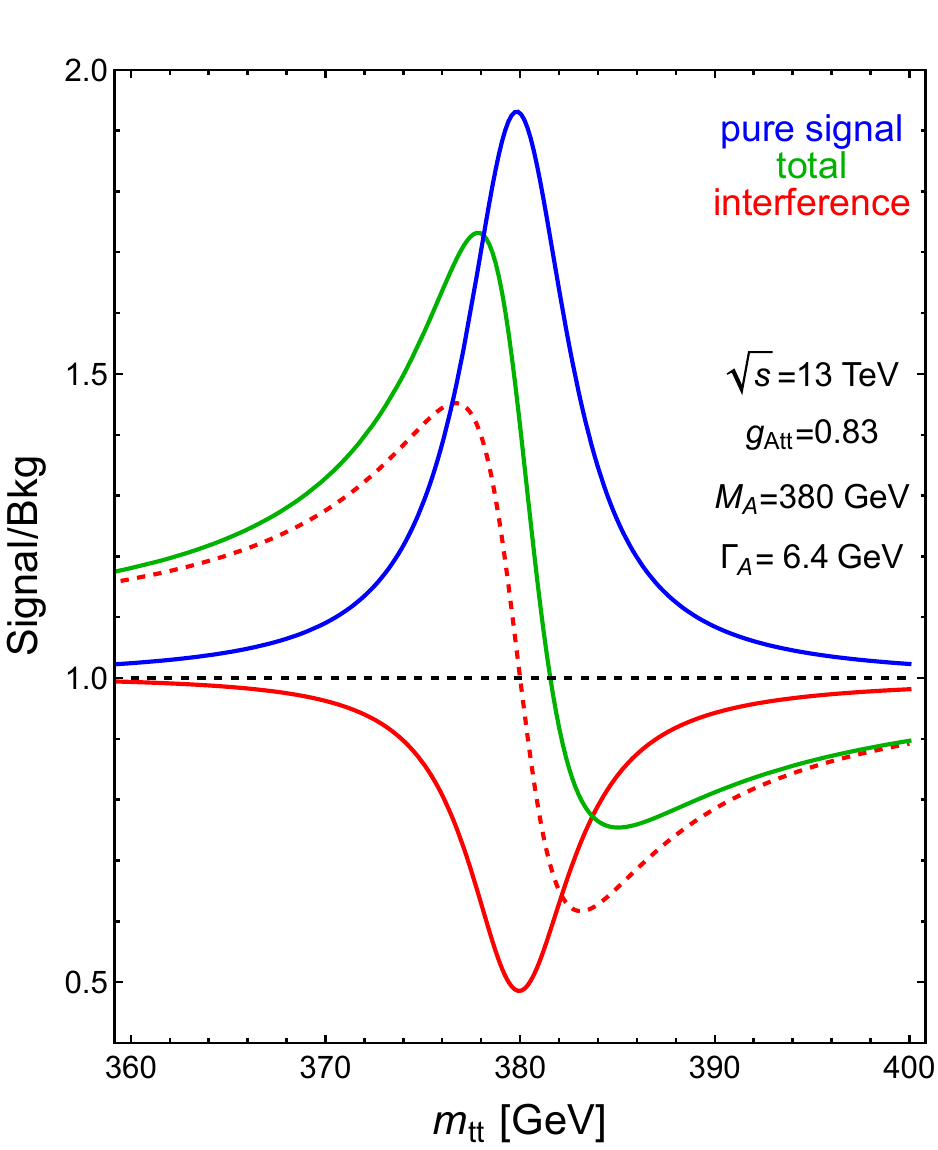} }
\end{center} 
\vspace*{-5mm}
\caption{\it The contributions to the line–shape of a pseudoscalar $A$ state with a mass (from left to right) $M_A\!=\!350, 365, 380$ GeV in the process $gg\! \to \! A \! \to \! t \bar t$ at the LHC with $\sqrt s=13$ TeV. We show the contributions from the pure signal only (blue lines), the real and imaginary interference contributions (dashed and solid red lines) and the total rates including the interference (green lines). The values of the $g_{Att}$ couplings, chosen to obtain the rate observed by CMS, and the resulting values of $\Gamma_A$, evaluated at leading order, are indicated.}
\label{Fig:interference}
\vspace*{-3mm}
\end{figure}

This can be seen in Fig.~\ref{Fig:interference},
where the impact of the interference effects on the cross section for the signal plus background, normalized to the cross section for the background alone, is shown as a function of the invariant mass of the $t\bar t$ system. Three values of the $A$ mass close to the $t\bar t$ threshold are chosen, namely $M_A=350, 365$ and $380$ GeV from left to right. In each case the coupling $g_{Att}$ is chosen so as to give an overall normalization that is compatible with the magnitude of the signal reported by CMS,  namely $\sigma(pp\to A\to t\bar t) =7.1$~pb~\cite{CMS}. The total width $\Gamma_A$ is completely fixed by this normalisation and is also indicated in the plots.~\footnote{ 
 {CMS mentioned in their analysis~\cite{CMS} several {\it ad hoc} values of $\Gamma_A$. For definiteness, they} {use in the figures and Appendix~C of~their paper $\Gamma_A^{\rm tot}/ M_A = 2$\%,  which corresponds to $\Gamma_A=7.3$ GeV for the main parameter set that they discuss, namely $M_A = 365$~GeV and $g_{t \bar t} = 0.78$ (the latter choice would lead to the  value $\tan \beta = 1.28$ in the 2HDM).} These total width  values do not correspond to the (leading order) ones that we use here, which are fixed for given $M_A$ and $g_{Att}$ values.}

As can be seen in Figs.~\ref{fig:365} and \ref{Fig:interference}, the real part of the interference (dashed red lines) changes sign across the nominal $M_A$ value, {becoming negative above $M_A$}, whereas the imaginary part (solid red lines) is always negative~\footnote{We note that both are suppressed as the $t \bar t$ threshold is approached from above because of the small size of the QCD background.}. Since the width is much smaller than $M_A$, the net effect of integrating the real part over the signal region is suppressed. Hence, the combined interference effect on the total cross section is negative and reduces the strength of the signal.  In principle, there is a dip in the $t\bar t$ mass distribution above the nominal mass and a greater sensitivity to interference effects could thus be obtained by comparing off–center bins. 
{However, all the interference effects for the total widths discussed here, $\Gamma_A \lsim 10$ GeV,} are smeared out once one takes into account the experimental resolution on the invariant $t\bar t$ invariant mass, which is much larger, $\sim 20$\%.

{
The effect of the interference should nevertheless be detectable for $A$ masses sufficiently far above the $2m_t$ threshold. Indeed, as can be seen from Fig.~\ref{fig:365}, contrary, e.g., to the case $M_A=350$ GeV shown in Fig.~\ref{Fig:interference} where the total width $\Gamma_A$ is rather small and the impact of the interference is restricted to a narrow window of a few GeV around the mass $M_A$,  in the case $M_A=365$ GeV the ratio between the  differential cross sections for the pure signal only and for the signal with the interference included is significantly smaller than unity for $m_{t\bar t}$ values a few tens of GeV to 100 GeV above the nominal $A$ mass.}~\footnote{This is similar to what occurs in the SM-Higgs case where it was noticed that in the production channel $gg\to ZZ \to 4f$, a large fraction of the Higgs–mediated cross section lies in the high–mass tail where the invariant mass of the $ZZ$ system is larger than $2M_Z$ \cite{H-width}. This allows a nice measurement of the Higgs total width $\Gamma_h$ despite its value.}

{
In fact, the interference is crucial to obtain the correct total cross section of $7.1$ pb reported by the CMS collaboration. This is exemplified in the left-hand side of Fig.~\ref{Fig:energydependence}, where we show the cross sections for the signal only and for the signal plus interference when integrated over the $t\bar t$ invariant mass, from the threshold value $m_{t\bar t}=345$ GeV to  a given $m_{t\bar t}^{\rm max}$ value. In the case of the signal only, an 
$m_{t\bar t}^{\rm max}$ value of about 10 GeV (corresponding to about $2 \Gamma_A$) above the nominal Higgs mass  is sufficient to obtain the total rate with enough accuracy. However, in the case of the signal plus interference, one has to integrate over  $m_{t\bar t}$ values up to 600 GeV to obtain the full rate.}

{A numerical example that is rather telling is that when integrating $m_{t\bar t}$ in the window [345 GeV, 375 GeV] one obtains for the  ratio (signal only)/(signal+interference) a value of about 1.2 while if the integration is performed in the much wider window [345 GeV, 600 GeV], on obtains a much larger ratio, $\approx 4$.}
{This huge difference in the rates should be measurable at the LHC, in particular if the resolution on the $t\bar t$ mass spectrum could be improved to about 10\%, which may be a realistic target.}
    
{Another feature of the $A$ scenario that might be of interest in the future is that the pure signal and the signal plus interference rates have different energy dependences.  This can be seen in the right-hand side of Fig.~\ref{Fig:energydependence} where the cross sections for the signal only and for the signal plus interference are shown as functions of $\sqrt s$, after integration over the full allowed $m_{t\bar t}$ range; the usual inputs for $M_A = 365$~GeV  and $g_{Att}=0.78$ are assumed. For example, one finds that the  ratio for signal-only to signal+interference increases by about 20\% between 13 and 30 TeV in the $pp$ centre of mass, and by a similar amount between 30 and 100 TeV.}
{We also show for comparison the energy dependence of $\eta_t$ production, taken from Ref.~\cite{Jiang:2024fyw}, which grows more rapidly.}

\begin{figure}[!ht]
\vspace*{-2mm}
\begin{center}  
\mbox{ 
\includegraphics[scale=0.35]{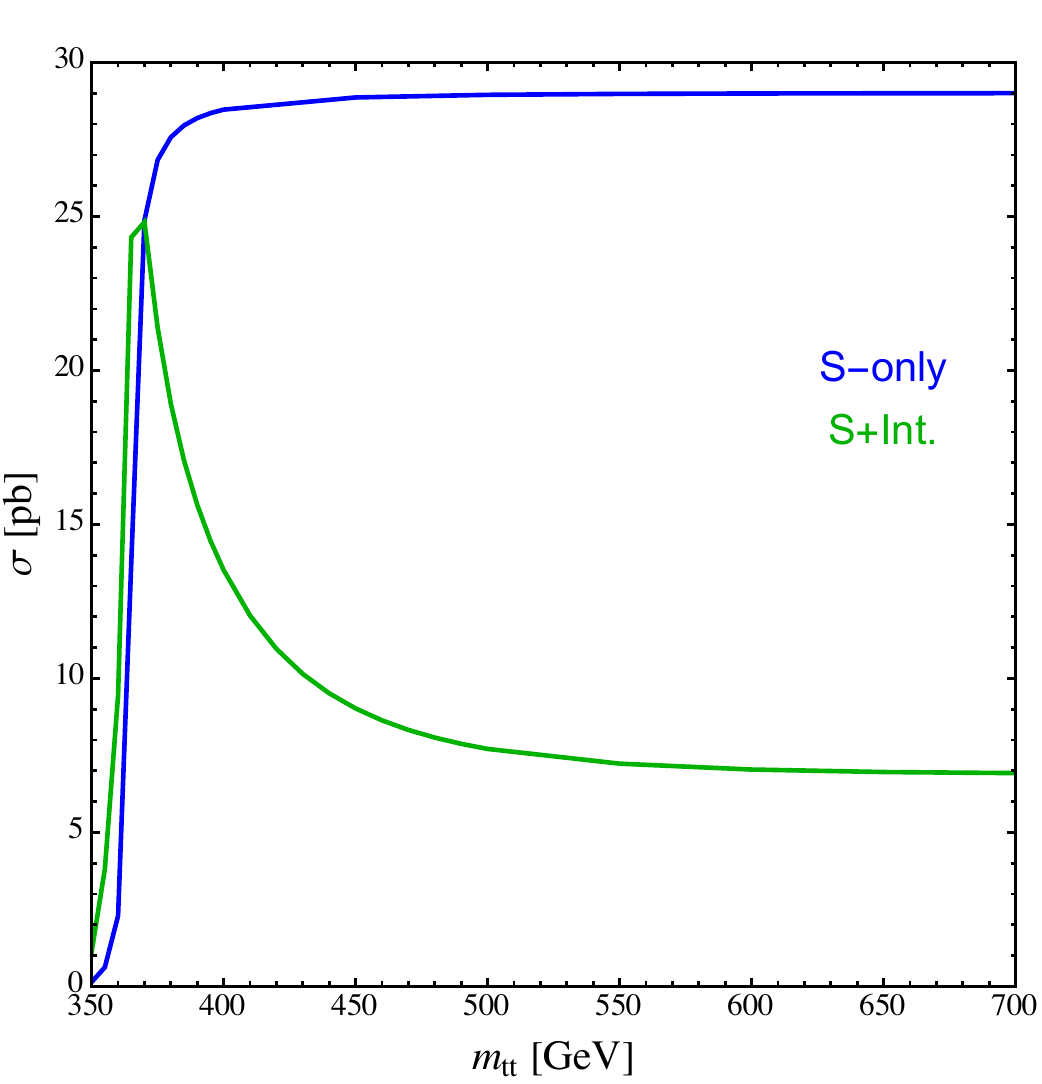}
\hspace*{-3mm}
\includegraphics[scale=0.35]{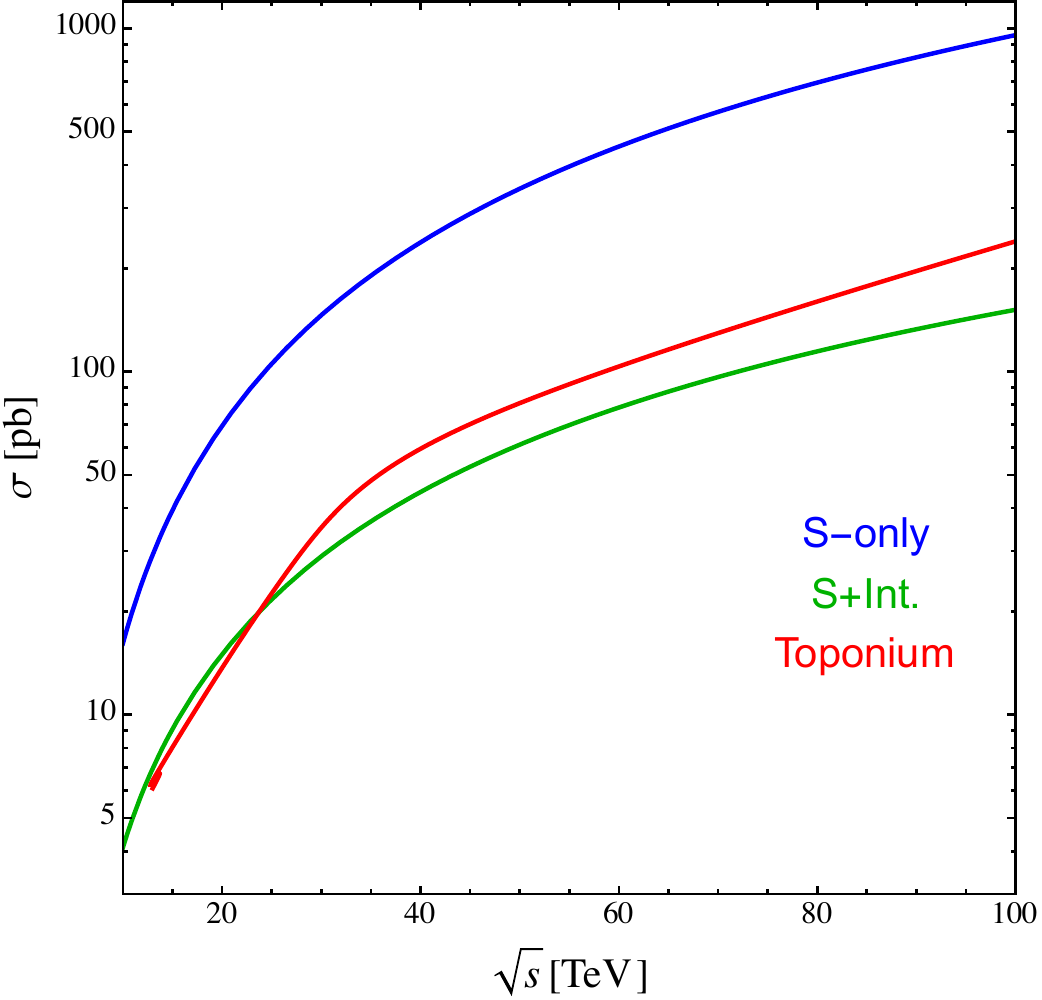} }
\end{center} 
\vspace*{-.8cm}
\caption{\it The production cross sections at the c.m. energy $\sqrt s = 13$~TeV for the $A$ signal only and for the signal plus interference in the process $gg\to A\to t\bar t$ for $M_A\!=\!365$ GeV and $g_{Att}=0.78$. Left: when integrated in the mass window $[2m_t, m_{t \bar t}$] for ranges $m_{t \bar t} \le 700$~GeV. Right: as functions of the c.m. energy $\sqrt s$ from 13 TeV to 100 TeV after integration over the full $m_{t\bar t}$ range. The energy dependence of $\eta_t$ production, taken from Ref.~\cite{Jiang:2024fyw}, is also shown for comparison.} 
\label{Fig:energydependence}
\vspace*{-1mm}
\end{figure}

{We conclude that in the effective scenario that we discuss here, while the peak-dip structure of the interference is difficult to observe at the LHC, as the total width of the $A$ boson is too small in general and is completely hidden in the much larger $t\bar t$ mass resolution,  there are other means that allow to probe the interference.  Another direct way to discriminate between the production of $\eta_t$ toponium and a pseudoscalar $A$ Higgs boson is discussed in the next Section.} 

\subsection*{4. Other production mechanisms for the A boson} 

Another aspect that shows a difference  between a toponium resonance and a pseudoscalar $A$ boson strongly coupling to top quarks would be the production of the latter in a completely different process. One such process is associated $A$ production with the lighter $h$ boson, $pp \to hA + X$. In our case, the only possible access to this final state is via gluon fusion, $gg\to hA$, through box diagrams in which virtual top quarks are exchanged~\footnote{In specific scenarios like, e.g., a general 2HDM, the dominant channel is the tree-level process in which a virtual $Z$ boson produced in $q\bar q$ annihilation yields $hA$ final states, $q\bar q \to Z^* \to hA$. However, both in our case and in the aligned 2HDM, the $ZhA$ coupling is zero and this this channel does not occur. For the same reason, the decay $A \to hZ$, which would be a good signature for an $A$ boson, is unobservable. In a general 2HDM,  bottom quarks can be exchanged in the box diagram but, as the $Abb$ couplings are not enhanced for the set of parameters that reproduce the the CMS signal, their contribution is small. This is also the case for $A$ production in association with $b\bar b$ pairs, which is discussed below.}.  

Another process of interest that would allow for a contrast in hadronic collisions is $A$ production in association with top quark pairs, $gg/q\bar q \to t\bar t A$. Although it is of lower order in perturbation theory compared to $gg\to hA$, as there is one electroweak coupling less at the amplitude level. However,  this process is affected by a less favorable phase space and its rates are suppressed at lower centre-of-mass energies like those of the LHC. 

We have calculated the cross sections for these two processes using the numerical programs {\tt HPAIR} and {\tt HQQ} of Ref.~\cite{Michael-web} along the same lines as in Ref.~\cite{Higgs-100TeV}, except that we include the NLO $K$-factor for the process $gg\to hA$, which increases the rate by a factor of two \cite{K-Hpair} (in Ref.~\cite{Higgs-100TeV} the calculation was performed in the 2HDM  where the $b$ coupling plays an important role but for which the NLO corrections were not known). For the $pp \to t\bar t A$ process, the QCD corrections are rather smaller \cite{K-Hqq}, and we ignore them. 

The hadronic production cross sections for the two processes are shown in Fig.~\ref{Fig:otherprocesses} as a function of the c.m. energy, with the inputs taken to be $M_A=365$ for the $A$ mass and the $g_{Att}$ couplings to top quarks that reproduce the CMS signal. At LHC energies, $\sqrt s=13$ TeV, one obtains relatively small cross sections for the two processes: $\approx 25$ fb for $hA$ and $\approx 15$ fb for $t\bar t A$.  With the much higher integrated luminosity that is expected to be collected in the future LHC upgrade, namely 3000 fb$^{-1}$, it might be possible to overcome such a small rate. At a 100 TeV collider, the rate is increased by two orders of magnitude (specifically, a factor of about 50 for $hA$ and about 200 for $t\bar t A$, assuming the same $A$ mass) and a possible detection becomes much more likely.  
     
\begin{figure}[!ht]
\vspace*{-1.3cm}
\begin{center}  
\includegraphics[scale=0.8]{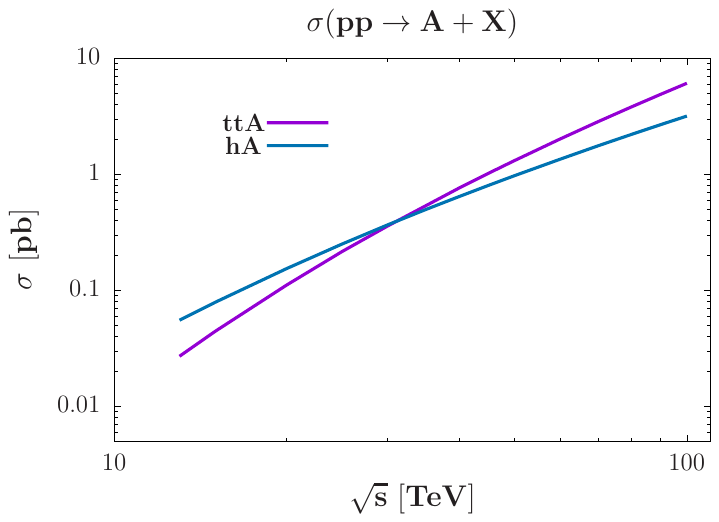} 
\end{center} 
\vspace*{-15.3cm}
\caption{\it The $A$ production cross sections via the processes $gg \to hA$ and $gg/q\bar q \to t\bar t A$ as functions of the $pp$ centre-of-mass energy $\sqrt s$, for the inputs $M_A\!=\!365$ GeV and  $g_{Att} =0.78$.}
\label{Fig:otherprocesses}
\vspace*{-4mm}
\end{figure}


In principle, the $A$ and $\eta_t$ scenarios could be differentiated via searches for subdominant $A$ decay channels like $A \!\to\! \gamma\gamma, Z\gamma, ZZ, W W $, which in some cases like $A \to \gamma\gamma$ and $A\to ZZ \to 4\ell$ with $\ell=e,\mu$ have a small experimental resolution to be sensitive to the interference effects. However, these channels should have very small branching fractions {as is expected also to be the case in the $\eta_t$ interpretation, as recently discussed in~\cite{Yang}, and they could in fact mimic those of toponium.} 
In other scenarios such as variants of the 2HDM, the decays $A\to b\bar b$ and $\tau^+\tau^-$ could lead, in some cases, into significant branching ratios that might allow for the channels to be observable.

Turning to $e^+ e^-$ colliders, as mentioned in the Introduction, there have been detailed studies of $^3S_1$ vector toponium in this case, see, e.g.,~Ref.~\cite{Reuter}, where the well-defined centre-of-mass energy would enable a narrow peak to be identified and measured. The spatial wave-function of the $^3S_1$ vector and $^1S_0$ pseudoscalar toponium states are expected to be very similar, so correlation of the $e^+ e^-$ signal with CMS signal would be convincing evidence that the latter is due to toponium. In turn, the direct production of an $A$ boson would be very difficult in this context since, in the  most favorable option,  associated production in  $e^+ e^- \to t \bar t A$, the expected cross section would be very small~\footnote{In the context of a 2HDM, there are other possibilities in $e^+ e^-$ collisions~\cite{Bechtle}. One is to look for associated production of the pseudoscalar $A$ with the lighter scalar $h$, which would require a centre-of-mass energy above $\approx 500$~GeV. However, since the $ZhA$ coupling is strongly suppressed in the alignment limit that should be considered, the cross section for $e^+ e^- \to h A$ would be suppressed.}. Another interesting possibility would be to run a linear $e^+e^-$ collider with an initial c.m. energy less than 500 GeV in the $\gamma\gamma$ mode and to produce the $A$ state as a resonance, $\gamma\gamma \to A$, again through triangular top quark loops; see, e.g.,~Ref.~\cite{Rohini} for a discussion.  

Finally, in realistic extended Higgs scenarios, the pseudoscalar $A$ boson is accompanied, in general, by other Higgs states. In the 2HDM for instance, there is an additional neutral CP-even $H$ and two charged $H^\pm$ states that are present in the spectrum (on top of the SM-like $h$ boson) and, in our context, should be heavier than $A$ as they have not been observed. These additional particles, if produced and observed, would give a strong weight to the $A$ hypothesis and even proof of its existence if these particles would also decay into $A$ bosons. Two examples of such a situation are the following. The  neutral $H$ boson can be produced in gluon-fusion  through top (and bottom) quark loops, $gg\to H$, and subsequently decay either into $t\bar t$ or into $ZA\! \to\! Zt\bar t$ final states; the latter mode is in fact not suppressed in the alignment limit and has a substantial branching ratio that would allow for its  observation \cite{Cheung}. Another possibility would be the production of the charged Higgs boson in the process $gb \to H^-t$ with the subsequent decays $H^-\! \to\! \bar t b$ or $H^-\! \to \! W^-A$, the latter possibly having a substantial rates especially in the alignment limit. 

One could also have associated production of $HA$ and $H^\pm A$ but, for the large $H,H^\pm$ masses required by present constraints, the rates for these electroweak processes are expected to be small even at higher-energy hadron colliders. Alternatively, at high-energy $e^+e^-$ colliders, the two channels  $e^+ e^- \to H A$ or $e^+ e^- \to H^+ H^-$,  which would require c.m. energies above respectively $\approx 1$~TeV and $\approx 2$~TeV in the scenario discussed here, would be possible and would provide reasonably large rates if kinematically accessible. 

\subsection*{5. Conclusions} 

The discovery of a pseudoscalar Higgs boson $A$ at the LHC would be a remarkable breakthrough into physics beyond the Standard Model, if confirmed. However, realistically, it is absolutely clear  that the most plausible interpretation of the CMS signal~\cite{CMS} would be that it is a $^1S_0$ quasi-bound $t \bar t$ state - pseudoscalar toponium, $\eta_t$. Discovering toponium 50 years after the November Revolution would be an unanticipated and welcome golden anniversary present for its charmonium cousin that was discovered {at SLAC and  Brookhaven} in 1974~\cite{J+psi}. 

Nevertheless, one should pursue the searches for additional Higgs bosons. In this paper, we  have highlighted the main differences between the production of a pseudoscalar $A$ boson and the toponium scenario, taking as an illustration the CMS signal. One possibility to contrast 
{
between the $A$ and $\eta_t$ possibilities would be to use the fact that in the former case the interference between the $gg\to A \to t\bar t$ signal and the $gg\to t\bar t$ QCD background is potentially important, particularly if $M_A$ is not very close to the $t \bar t$ threshold. 
As we have also shown in this paper, other production processes, eventually at high-energy proton or future lepton colliders, and complementary measurements at these colliders, also allow to determine completely the profile of a pseudoscalar Higgs resonance and probe fully the new physics scenario that lies behind it. 

In the mean time we look forward to experimental clarification of the CMS signal - in particular whether it is confirmed by the ATLAS Collaboration - as well as theoretical work to refine our understanding of the properties of the putative $\eta_t$ state - aiming in particular to reach consensus on its conceptual interpretation.\bigskip

\noindent {\bf Acknowledgements}:  
We thank Christian Schwanenberger for discussions of the CMS data and Giorgio Arcadi for discussions on the 2HDM. The work of J.E. was supported by the United Kingdom STFC ST/T000759/1 and the one of A.D. by the grant PID2021-128396NB-I00.

\end{document}